# Deepfake pornography as a male gaze on fan culture

Inna Suvorova, University College of London, IOE

**Abstract:** This essay shows the impact of deepfake technology on fan culture. The innovative technology provided the male audience with an instrument to express its ideas and plots. Which subsequently led to the rise of deepfake pornography. It is often seen as a part of celebrity studies; however, the essay shows that it could also be considered a type of fanfic and a product of participatory culture, sharing community origin, exploitation by commercial companies and deep sexualisation. These two branches of fanfic evolution can be connected via the genre of machinima pornography. Textual fanfics are mainly created by females for females, depicting males; otherwise, deepfake pornography and machinima are made by males and for males targeting females.

Deepfake is an interesting but dangerous phenomenon that can undermine faith in the authenticity of video information. The most scandalous deepfakes are of a political kind; however, most deepfake videos are deepfake pornography that targets and harms mainly females from the entertainment sector.

As deepfake pornography mostly depicts famous persons, it is often considered as a part of celebrity studies. This type of video can be compared with leaked private photos/videos with celebrities. The well-known case of that kind is high jacking nude photos of Scarlet Johansson from her mail account, by Christopher Chaney, who got 10 in prison for that in 2012 (Wired, 2012).

However, many similarities link deepfake pornography with fan culture and other related topics as machinima pornography. That could be demonstrated on the example from the game industry: Cyberpunk 2077 players created a mod that allows players to have in-game sex with Johnny Silverstone (the character with the appearance of Keanu Reeves). It was achieved by transferring the character's appearance to NPC (The Verge, 2021).

The purpose of this essay is to show that deepfake pornography can also be considered as a part of fanfiction along with machinima and machinima pornography. It can become a new chapter in fan movement, not of a pornography kind.

This essay will start by scrutinising the phenomenon of deepfake and deepfake pornography in Part 1, leading to the Part 2 where we will have a closer look at the similarities and differences between deepfake pornography and fan cultures.

**Deepfake as a technology, its history and existing types**

Deepfakes as a phenomenon arose in 2017 and since then draws attention as a grim technology that affects our perception of the world. Latest deepfake that shook British society is the Her Majesty the Queen's speech made by Channel 4 (BBC, 2020). Most of the articles about deepfakes are dedicated to the way researchers can detect whether the video or photo was manipulated or not.

Sensity – the company that works on video threats – identified 49,081 deepfake videos in June 2020, the number increased by 330% since July 2019 (2020).

Wikipedia provides us with the following definition: deepfakes (a portmanteau of "deep learning" and "fake") are synthetic media in which a person in an existing image or video is replaced with someone else's likeness. However, reality provides us with mismatching examples. On the deepfake video with Speaker of the United States House of Representatives Nancy Pelosi, we can still see Mrs Pelosi, where the original video was made. The doctored part is her way of speaking and face expression. The deepfake creators made her look drunk (2019).

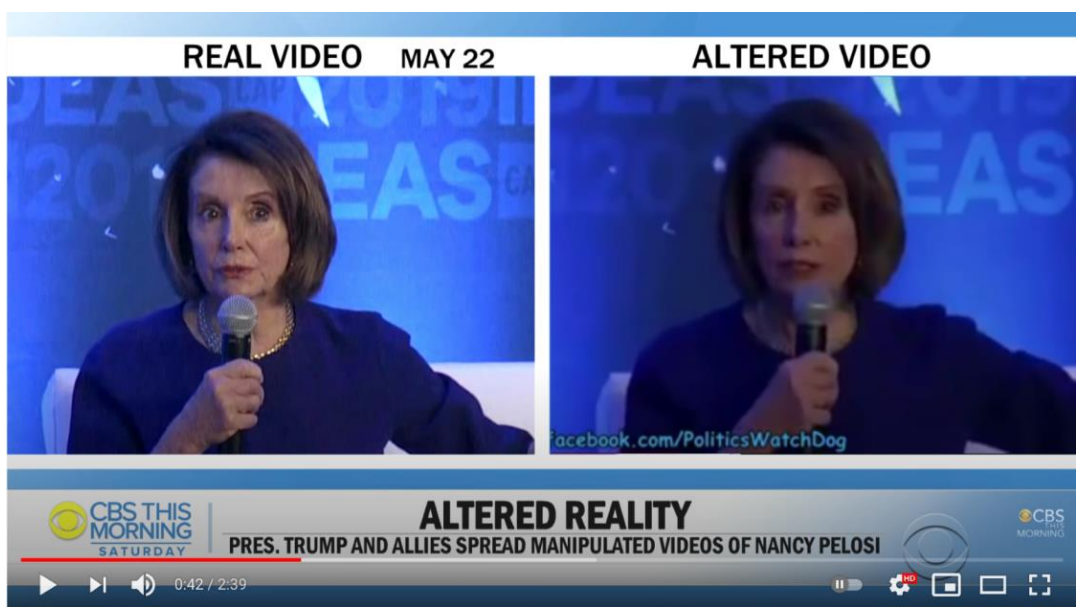

Image 1. Nancy Pelosi, real and altered video

So, the definition that Microsoft (2020) provides in its blog post about the release of the new deepfake detection software Video Authenticator is more precise: "deepfakes are synthesised media that are photos, videos or audio files manipulated by artificial intelligence (AI) in such a way that the fake is hard to detect". These videos are a type of AI-generated video, many of which are not fake and serve hugely different purposes.

Deepfakes were made possible by the development of several technologies. In 2002 Lev Manovich wrote, "And even for ILM (programme) photorealistic simulation of human beings, the ultimate goal of computer animation, still remains impossible" (2002, p. 182). However, 14 years later in 2016, it became possible.

Stanford University presented a program called Face2Face (2016), which could analyse one person's facial expression in real-time and transfer it to another.

In 2017, researchers from the University of Washington trained a neural network to generate videos of former US President Barack Obama. The project "Synthesizing Obama" (Suwajanakorn, S., Seitz, S.M. and Kemelmacher-Shlizerman, I., 2017) demonstrated how to "make" a video with Barack Obama's voice to make any speech, with the lips articulating the right words.

Researchers at the University of Alabama at Birmingham (USA) Mukhopadhyay D., Shirvanian M., Saxena N. (2015) shows how easily the person's specific speech can be generated using a 3–5-minute sample of speech from for example YouTube. The quality of AI-generated speech is high, and it will pass most of the tests.

Altogether these three technologies, voice generating, transfer one's facial expressions and mimics to another face and solving the problem of lip's articulation along with enormous price reduction and simplifying access to computing power made deepfake as a massively available technology possible.

**Deepfake types**

All the deepfakes can be divided into three very unequal "zones" relying on the legal origin and exposure of the fact that the video was made using artificial intelligence. It is a mimic genre that blends news, epochs, politicians, and celebrities and adapts to the wide variety of video production needs. It is an excellent example of media convergence, which is "more than simply a technological shift. Convergence alters the relationship between existing technologies, industries, markets, genres, and audiences. Convergence alters the logic by which media industries operate and by which media consumers process news and entertainment". (Jenkins 2006, 15-16).

White zone. The case when the deepfake provenance is exposed and all the participants agree to take part. We can refer to this type all the videos made by the producers who have legal rights on using images and rare cases when people provided their faces for making deepfakes on purpose. In the film "Welcome to Chechnya" (2020) European activists provided their faces to protect the identities of the people who fled from Chechnya. The latter are persecuted by their compatriots for their sexual orientation. Using deepfake technology allows keeping face expressions and voice intonation, however, hides the identity.

Deepfakes are adopted by the video production industry. Luke Skywalker's appearance in the series Mandalorian was possible because of the transfer of Mark Hamill's face in his younger age to the Max Lloyd-Jones face, who played this role. In this case, Mark Hamil was aware of using his face and even voiced the character. The actors agreed to the use of technology, the audience knew they were watching a feature film. Effects of that type are not something new for the industry, but deepfakes made it cheaper:

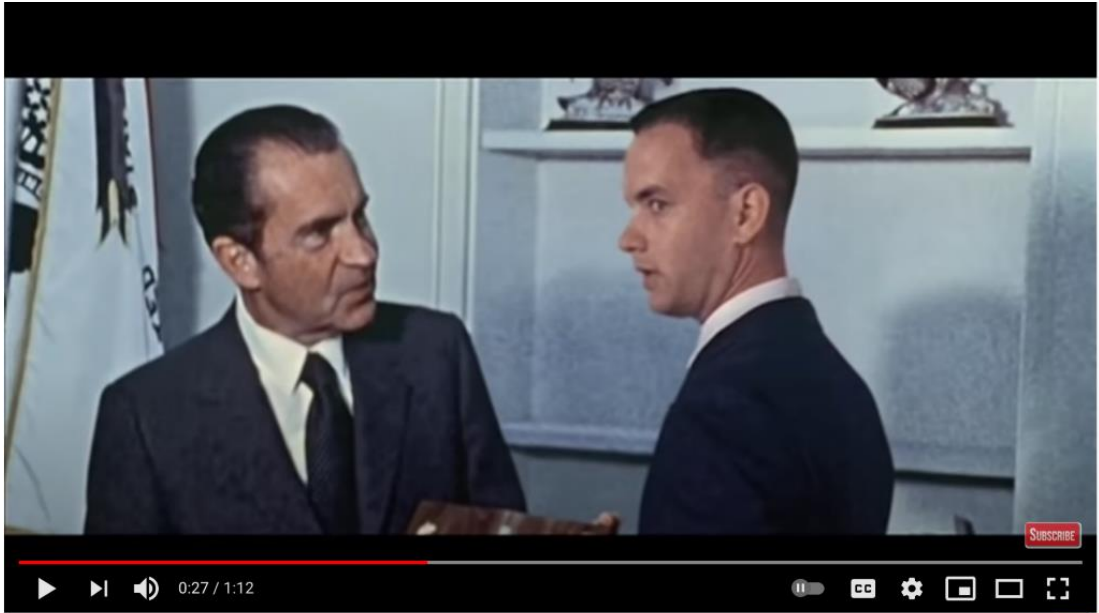

Image 2: Forrest Gump meats President Nixon (Forrest Gump, 1994).

Grey zone. Posthumous deepfakes are controversial. There is no procedure for obtaining consent from the person who already died. In the media campaign #stillspeakingup the media agency Publicis shows the deepfake resurrection of murdered Mexican journalists that was undertaken to uncover the tragic events around the murders of journalists.

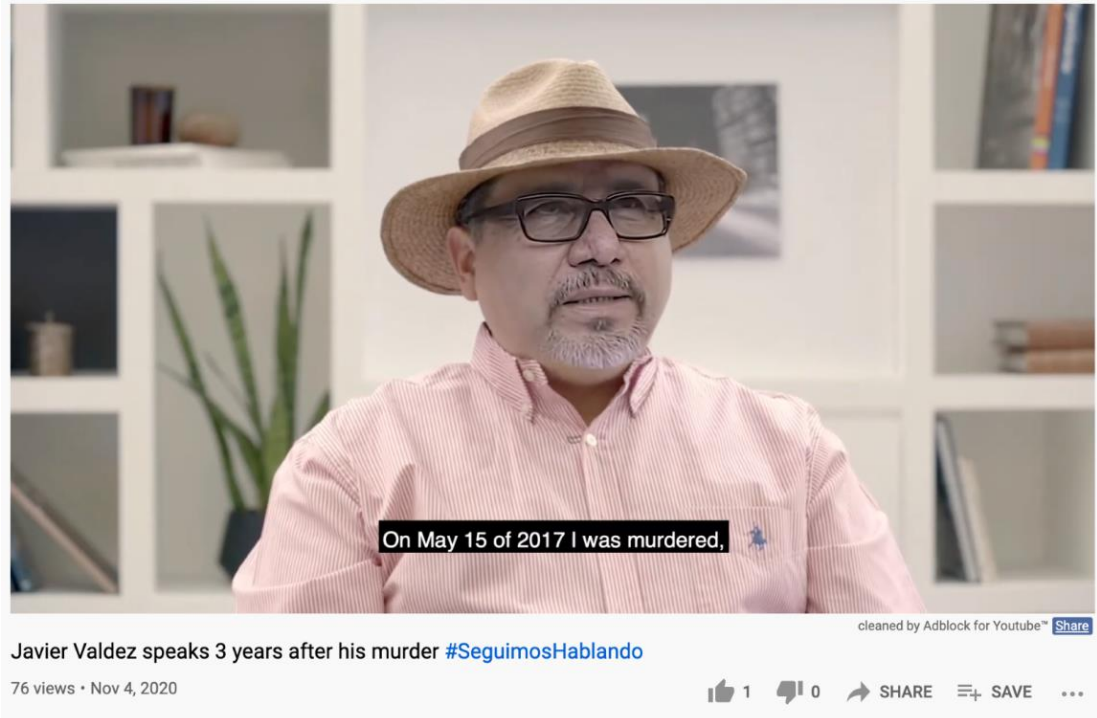

Image 3: Deepfake video with murdered in 2017 journalist Javier Valdez (2020).

There is even more controversy in the usage of deepfakes in the Unfinished votes campaign (2020) in the United States that used AI-generated video of the teenager Joaquin Oliver killed in a school shooting. He calls for participation in elections on the video.

There is some number of entertaining videos, where deepfake technology is used. Already mentioned in this essay video "Deepfake queen to deliver Channel 4 Christmas message" (2020) is a controversial act; the usage of the deepfake technology is exposed, but an agreement for using person's appearance was not given.

Black zone. Deepfake video can be made for political or social (or both) disruption, like this video made by Extinction Rebellion Belgium where Sophie Wilmès, Deputy Prime Minister of Belgium, claims ecological problems as the main reason of the coronavirus epidemic (Galindo, 2020).

Another worrying case is the using of deepfakes for cyberbullying. Indian journalist Rana Ayyub became a revenge campaign target, including a deepfake pornography video with her images, for her investigative reports (Washington Post, 2018).

At this point, it is essential to remember Nico Carpenter's (2007, 105-122) idea of underestimation of "old" media and their significant role in modern days. Fake news and deepfake circulation can be if not stopped but deterred by traditional practices of fact-checking. During the coronavirus epidemic, the amount of circulating fake information was so high that some people returned to the traditional media as a news source (Whitebloom, 2020).

Nevertheless, the most significant share of the deepfake black zone and deepfake at all is non-consensual AI-generated pornography or deepfake pornography, that accounts for 96% of all the deepfakes, according to the Sensity's report (2019).

On this point, we should scrutinize the phenomenon of deepfake pornography, when porn actors' faces are replaced by AI with the faces of other people, e.g., actresses, celebrities.

Controversial nature of deepfake pornography previously led to the decision of most of the video hosting platforms to eradicate such content. As Christian Fuchs mentions, "corporate platforms owned by Facebook, Google and other large companies strongly mediate the cultural expressions of Internet users" (2014, 56). So, most of the big media platforms prohibit their users from posting any pornographic content. Deepfake pornographic videos banned by Twitter (the only big platform that allows pornographic content to some point). Even Pornhub, the most famous resource for hosting pornography, banned AI-generated porn as non-consensual (Cole, Vice, 2018).

So now the deepfake pornography resources are mainly hosted on biggest porn websites like xvideos.com and xnxx.com. They both are more popular than Pornhub keeping 8 and 9 places in the list of most popular websites in the world according to the SimilarWeb (2020).

After explaining what deepfake pornography is, we transfer to the features that deepfake pornography shares with fanfiction.

**Deepfake pornography and fan texts. Similarities and differences**

Before diving deeper into controversial relationships between fan texts and deepfake pornography, which as we suppose could be a part of it, we should clarify the way the term "text" is used in this essay. Fan culture is represented by different genres, written texts, videos, events. Here we use term "text" not only as something written but, according to Andrew Burn and David Parker, "what we recognise as a 'media text' still includes, importantly, film and television, it is subject to the same processes of rapid change which all kinds of text are experiencing in the age of digital media" (2003, 1).

We will start from the features that fanfiction and deepfake pornography have in common.

**Similarities**

A good example we can find in the American noir psychological thriller "Vertigo" by Alfred Hitchcock, released in 1958. The main character of the movie is John "Scottie" Ferguson. He describes himself as "a man of independent means" and subsequently transforms his girlfriend's appearance to the image of a young lady she played before, that he fell in love with.

Modern fan culture and deepfake pornography sometimes resemble "Scottie". Authors use their power to force the character to behave the way they want under circumstances they find appropriate and engaging.

Both deepfake pornography and fanfiction creation take power to spread ideas, plots, and meanings from the official media and give them to the amateur hands for good or evil. Fanfiction authors do not expect favours from the creators of beloved characters; they create themselves.

Unlike the celebrity culture where fans can only indirectly affect celebs' behaviour and choice via shame or approval, both deepfakes and fanfiction of any type allow creators to manipulate characters the way they want.

**Participatory culture**

There is a point that links fanfiction and deepfake pornography. Both have their origin in participatory culture.

Henry Jenkins (2009, 5-6) highlights five main traits of participatory culture:

> "1. relatively low barriers to artistic expression and civic engagement,
>
> 2. strong support for creating and sharing creations with others,
>
> 3. some type of informal mentorship whereby what is known by the most experienced is passed along to novices,
>
> 4. members who believe that their contributions matter, and
>
> 5. members who feel some degree of social connection with one another (at the least, they care what other people think about what they have created)."

Fanfiction creation is not something new, first texts that can be called "fanfics" were related to the Sherlock Holmes book series and appeared soon after Conan Doyle "killed" his character to dedicate his time to serious historical texts (Wired, 2009). The Baker Street Irregulars literary society that "is dedicated to the study of Sherlock Holmes, Dr. Watson, Sir Arthur Conan Doyle, and the Victorian world" was founded in 1934. The development of the Internet boosted fanfiction culture, allowing authors from all over the world to share, create, argue and support each other.

Deepfake pornography is much younger; technologies made possible this type of video creation by non-professionals only in 2017. However, it is also a participatory culture product. Its history starts at the beginning of 2018 from a community on Reddit of about 100,000 users, who shared ideas and results. Later the community was deleted by Reddit administration. On January 2018 Vice published an article based on discoveries made in this community called "We Are Truly Fucked: Everyone Is Making AI-Generated Fake Porn Now" (2018).

As Jose Van Dijck mentions (2009, 42-43) "the result is a participatory culture which increasingly demands room for ordinary citizens to wield media technologies – technologies that were once the privilege of capital intensive industries – to express themselves and distribute those creations as they seem fit".

Both deepfake pornography and fan culture as we know it was made possible by modern technologies.

**The dark side of participation is exploitation.**

This shift from participatory culture as something refectory and secondary to more independent and meaning making was also noticed by Jenkins et al (2013, 36), who gave a positive evaluation to the growing level of interaction between fanfiction authors and official producers:

> "All of this suggests ways we are revising the concept of participatory culture to reflect the realities of a dramatically altered and still-evolving mediascape. We are moving from an initial focus on fandom as a particular subculture to a larger model that accounts for many groups that are gaining greater communicative capacity within a networked culture and toward a context where niche cultural production is increasingly influencing the shape and direction of mainstream media. We are moving from focusing on the oppositional relationship between fans and producers as a form of cultural resistance to understanding those roles as increasingly and complexly intertwined".

Opposing to Henry Jenkins, Christian Fuchs and Marisol Sandoval pay significant attention to the commercial, capitalist side of the participation culture, where producers and media companies benefit from non-commercial content-creators. They write:

> "We have argued that the discourse on alternative and participatory media should be situated within the context of the analysis of capitalism. Capitalism brings about structural inequalities that shape the limits and potentials of alternative media projects. Power relations and the unequal distribution of resources stratify public visibility of actors and opinions. Giving people a voice by involving them in media production does therefore not mean that their voice is also heard. Participatory production processes can also be used for advancing repressive purposes and profit accumulation". (2010, 148)

Fan-produced texts are often victims of reversed "poaching" when corporations "borrow" ideas through online competitions, reading forums, collecting feedback, stealing designs, and use fans as a "quality control" and as a primary consumer of media products and related goods.

Both fan culture and deepfake pornography serve as an instrument of profit by the people who are not engaged in their creation. Deepfake pornography is the popular content on pornographic web sites. Its creators will hardly claim copyright to their works to get royalties.

Now we will scrutinise one more similar trait of fanfiction and deepfake pornography – interest to the relationships and sex.

## Sexualisation

Everything related to porn/sex/erotic bear a social stigma. Susan Sontag in her essay "The pornographic imagination" writes that "a society so hypocritically and repressively constructed that it must inevitably produce an effusion of pornography as both its logical expression and its subversive, demonic antidote" (1969, 207).

Fanfiction as a form of literature not only full of imaginary worlds, magic and spaceships and simultaneously is deeply imbued with themes of relationships, sex, pairing, erotic and textual pornography. Susan Sontag rightly observes that there is no difference in imagining and writing about "teeming planets" and spaceships, or "duration of orgasms", "the size of organs", "amount of sexual energy". Both do not exist. (Sontag 1969, 213).

Just a quick look on a popular fanfiction website https://www.fanfiction.net/ shows that the search on the website is constructed in the way to help the user find pairs "character A" and "character B", allows user to find the texts with "pairing" (where author construct the relationship between chosen character), and define age restrictions from K – suitable for kids to M – mature content for adults. MA, the most "mature" category, was exiled to another website.

Sometimes it seems that film producers intentionally get rid of most relationship forms in their production. That allows to partly avoid age restrictions that exist in many countries and ignite fans' imagination. That led to the bromance boom started in 2005 (DeAngelis 2014, 1) and has not ended until now.

"The Triple-A Engine theory posits that access, affordability, and anonymity serve as the three primary factors that drive increasing OSA" (Online sexual activity) (Larsen, K., Zubernis, L. 2012, 57).

Some of the fanfiction texts are brilliant; some are raw amateur crafts. According to Sontag "to put it very generally: art (and art-making) is a form of consciousness; the materials of art are the variety of forms of consciousness. By no aesthetic principle can this notion of the materials of art be constructed as excluding even the extreme forms of

consciousness that transcend social personality or psychological individuality" (Sontag 1969, 212)

That is how we come to the idea from the essay introduction. Deepfake pornography is often connected with celebrity studies, which could be a simplified way of understanding it. However, the celebrities' voices are clearly heard when somebody violated their personal borders by using their image for producing pornography.

The fact that the deepfake pornography is less about celebrity studies and more about fan culture was highlighted by Milena Popova in the article "Reading out of context: pornographic deepfakes, celebrity and intimacy" (2020):

> "I use secondary literature and my own past original research to compare deepfakes to three other types of sexualised audience engagements with celebrity: nude hacks (private, intimate images of celebrities obtained illicitly and shared beyond their original intended audience); Real Person(a) Fiction (RPF – a subset of fanfiction, frequently erotic, that fictionalises celebrities); and slash manips (still images created through digitally combining pornographic images with celebrity faces, created in communities that overlap with or are adjacent to RPF and wider fanfiction spaces)".

However, the missing piece in this puzzle, which expands our understanding of AI-synthesised pornography from celebrity studies to fan studies is machinima pornography. The phenomenon of machinima helps us to find even more profound interconnections between fan culture and deepfake pornography.

## Machinima and machinima pornography

Machinima is the term that describes the usage of real-time computer graphics engines to create a cinematic production. It is significantly older than deepfakes, the first video of that kind, "Diary of a Camper", based on the Quake engine, was released in 1996.

Since 1996 computer and video games have become much more beautiful and realistic. The game industry, literature and cinema production are often interconnected, like in the Witcher Universe case. There are the book series by Andrzej Sapkowski, the game, Film adaptation, TV series by Netflix, and a massive fanfiction on the topic, including videos that are made from pieces of games and movies by the sophisticated assembly.

Mizuko (Mimi) Ito describes (2011, 51) machinima as:

> "part of a proliferating family of emerging DIY fanvids, remixes, and parodies that we see spawning on the internet today. Like many other DIY video forms, machinima grows out of the creative energies of popular culture enthusiasts, fuelled by increasingly accessible digital media production tools and online video distribution. Although machinima's roots in gaming platforms and cultures make it unique, it shares a history and future with other kinds of DIY fanvids".

Some examples we can find blend different types of texts. Returning to the example from the introduction with Johnny Silverstone mod in Cyberpunk 2077, we see in it all the four types of texts we discussed:

1. Fan culture (embedding own narrative into the story, pre-created by someone else).
2. Machinima pornography (using a game engine for making pornography).
3. Deepfake pornography (using someone else's appearance to make porno).
4. Celebrity studies (Johnny Silverstone has Keanu Reeves' appearance).

Being quite demanding in production machinima pornography is often financed by crowdfunding and made by professionals. That is the current way for fans to get what they want. There are video production studios like Lewd Gamer, Naughty Machinima and Studio FOW, which specialised on producing such type of content (Madden 2017).

Studio FOW is famous for crowdfunded "adult" game "Subverse"; however, they also created a porno movie "Lara in Trouble" (based on the game "Tomb Raider", not on the film with Angelina Jolie, but with the character which shares the general outlook), and pornographic videos based on video games like BioShock Infinite (Bioshag: Trinity – Studio FOW's video), Dragon Age II (Siren's Call - Studio FOW's video). However, their idea of making porn video on Overwatch was prohibited by the game owner Blizzard.

Machinima pornography is not a spontaneous or isolated incident; research undertaken in the game Second Life by Débora Krischke Leitão, Laura Graziela Gomes and Raira Bohrer dos Santos (2016) shows that there is a community around machinima pornography based on the game, with special reward and even a journal on that topic.

**Differences**

Deepfake pornography and fanfiction texts have many common traits. However, we cannot ignore the difference between them. There are two notable differences – public attitude and audience that produce and consume them. The first point is strong enough, but the second one is not that obvious as it can seem on first sight.

**Public attitude**

The first difference is the way of producing and consuming fanfiction, and deepfake pornography is treated in society. It is reasonable to discuss in the way that was offered by Pierre Bourdieu. According to him (1977, 164):

> "This experience we shall call doxa, so as to distinguish it from an orthodox or heterodox belief implying awareness and recognition of the possibility of different or antagonistic beliefs. Schemes of thought and perception can produce the objectivity that they do produce only by producing misrecognition of the limits of the cognition that they make possible, thereby founding immediate adherence, in the doxic mode, to the world of tradition experienced as a "natural world" and taken for granted".

Now, after more than a hundred years of fanfiction development, creating fanfics of any kind and enjoying reading them is considered as a norm by our society. Famous "The Fifty Shades" erotic trilogy was a Twilight fanfiction series in the beginning.

Deepfakes of any kind are new to our society and threatening. Moreover, there are no reliable ways of their detection without special tools. The society is terrified by possible consequences of using deepfakes, political statements that have never been told by politicians or revenge porn that was created artificially and never took place. It is hard to see positive sides of technology which makes us too vulnerable to accept it even as heterodox.

**Fandom, deepfake pornography and machinima porn audience**

The tremendous difference between fanfiction, deepfake pornography and machinima pornography is in the audience gender structure. Audiences who create and who consume are strikingly different in these cases. First one is made by females and for females, the second and third ones are made for males and by males.

"Fandom scholars largely acknowledge the online community as a female-dominated space, and one of the primary finish activities in which females engage in the reading and writing of fanfiction". (Larsen, K., Zubernis, L 2012, 58).

Surprisingly, TOP-10 popular pairings are Male/Male.

| No. | Pairing | Canon | Type | Race | Works | Change |
|---|---|---|---|---|---|---|
| 1 | Castiel/Dean Winchester | Supernatural | M/M | White/White | 84563 | - |
| 2 | Sherlock Holmes/John Watson | Sherlock (BBC) | M/M | White/White | 61544 | - |
| 3 | Derek Hale/Stiles Stilinski | Teen Wolf | M/M | White/White | 59286 | - |
| 4 | Bucky Barnes/Steve Rogers | The Avengers | M/M | White/White | 49659 | - |
| 5 | Draco Malfoy/Harry Potter | Harry Potter | M/M | White/White | 40312 | - |
| 6 | Steve Rogers/Tony Stark | The Avengers | M/M | White/White | 36264 | +1 |
| 7 | Harry Styles/Louis Tomlinson | One Direction | M/M | White/White | 33126 | -1 |
| 8 | Keith/Lance | Voltron: Legendary Defender | M/M | Ambig/POC | 30317 | - |
| 9 | Aziraphale/Crowley | Good Omens | M/M | White/White | 28208 | +50 |
| 10 | Dean Winchester/Sam Winchester | Supernatural | M/M | White/White | 27696 | -1 |
| 11 | Magnus Bane/Alec Lightwood | Shadowhunters | M/M | White/POC | 24787 | - |
| 12 | Katsuki Yuuri/Victor Nikiforov | Yuri!!! on Ice | M/M | White/POC | 24333 | -2 |
| 13 | Jeon Jungkook/Kim Taehyung | BTS | M/M | POC/POC | 23339 | - |
| 14 | Rey/Kylo Ren | Star Wars | F/M | White/White | 21306 | +15 |

Image 4. Screenshot with the Top 20 pairings on AO3. Made February 11, 2021

The illustration shows statistics on the most popular pairings from AO3 website that hosts fanfiction. It is shown up to 14 places to highlight the position where females first appear in the chart.

Deepfake pornography is entirely different. As Henry Ajder et al writes (2019): "deepfake pornography is a phenomenon that exclusively targets and harms women. In contrast, the non-pornographic deepfake videos we analysed on YouTube contained a majority of male subjects". He adds: "all but 1% of the subjects featured in deepfake pornography videos were actresses and musicians working in the entertainment sector. However, subjects featuring in YouTube deepfake videos came from a more diverse range of professions, notably including politicians and corporate figures".

A research data provided by Sensity (2020) shows that 99% of the females whose images were non-consensually used for deepfake creation are actresses and celebrities working in the entertainment sector. The 1% - are women from the news and media sector.

Why are women affected that much? We should probably search for the answer in our culture.

Critic and art historian Abigail Solomon-Godeau writes:

> "Somewhere between the last decades of the eighteenth century and the end of the third decade of the nineteenth, the heroic male nude, alpha and omega of French academic art practice and foundation stone of history painting - the most exalted and prestigious form in the hierarchy of genres - lost its privileged position in practice, to be increasingly eclipsed by the female nude" (1993, p. 286).

Since then, the majority of the research on nudity is concentrated on female nudity as deepfake pornography is. "Women are depicted in a quite different way from men – not because the feminine is different from the masculine – but because the 'ideal' spectator is always assumed to be male, and the image of the woman is designed to flatter him." (Berger 2008, 99)

This explains to us why deepfake pornography mostly targets women. Because "from earliest childhood she has been taught and persuaded to survey herself continually. And so she comes to consider the surveyor and the surveyed within her as the two constituent yet always distinct elements of her identity as a woman." (Berger 2008, 68) Much work has been made to rid a woman of this attitude. Moreover, we should remember that the Barger's book was published in 1972. However, these patterns are still alive.

Who produce deepfakes? "There are at least four major types of deepfake producers: 1) communities of deepfake hobbyists, 2) political players such as foreign governments, and various activists, 3) other malevolent actors such as fraudsters, and 4) legitimate actors, such as television companies." (Westerlund 2019). But when we speak about deepfake pornography, only two points are suitable: communities (as we mentioned before deepfake started from the Reddit community) and malevolent actors. In case of, for example, revenge porn. And "like most pornographic content, it is predominantly produced by and for a male audience, although this time (fictionally) starring women who have not given their consent". (Öhman 2020)

Thus, deepfake porn audience, targets and producers are fanfiction audience, character and producers, but inside out.

Probably, when more females adopt the deepfake making technology, we will see significant changes in the persons who will be featured in these videos and in the audience that watches such videos.

What about the transitional genre of machinima? As Darlene Rose Hamptons mentions (2010), unlike fanvid, relatively feminine genre, machinima is a male-dominated space. There is no doubt that machinima is a part of fan culture, an example of fan culture segment predominated by men. Thus, we cannot give up the idea of considering deepfake pornography as not a part of fanfiction just because its producers and consumers are male.

Maybe it is a fanfiction too. Just for boys.

**Conclusion**

Henry Jenkins (2006, 131) mentions that "fans have always been early adopters of new media technologies; their fascination with fictional universes often inspires new forms of cultural production, ranging from costumes to fanzines and, now, digital cinema".

A significant amount of deepfake pornography and machinima pornography could be attributed as a part of fan culture. Its existence is the natural way of technology development. We can assume this way of video creation will be adopted by the official media to a more considerable extent.

There is an option that, like in the case of the antiviral software, deepfake detection software will be a step behind deepfake producers. However, there is still a chance that soon the technology which help to define deepfakes will emerge.

As it was many times in our history, humanity was first terrified with new technologies that later were adopted and accepted as practical or profitable. The same way deepfakes will find its niche.

Deepfake production is at the beginning of the arousal; it will flourish both in commercial video production and non-commercial. Creators of deepfake pornography choose the most obvious and less laborious way of getting content they want. However, the deepfake technology itself armours fan culture with the reliable instrument of creating a type of content they want, staging stories written before or generating their meanings. Moreover, yes, finally the erotic scenes will come back on screens long after the age restriction censorship and producers deprived the audience of such plot twists and turns.

Legal issues connected with deepfake pornography have arisen and will grow in future. While working on this essay, the author faced a situation when her close relative had an uncomfortable experience of becoming a victim of a semi-friendly prank of making a (non-pornography) deepfake with him. AI-generated nature of the deepfake video was evident; however, it did not make the experience pleasant. It raises a question of protection from the usage of personal photo and video records for creating deepfakes and responsibility for making it.

It is necessary to raise awareness about the phenomenon of deepfake. As people should realise that how truthful and convincing video could look, there is a possibility that it was AI-manipulated. That gives victims of deepfake at least the ability to explain what happened to them and be understood by others.

The proper attribution of the deepfake video will help build a base for legal norms. Now we see a growing number of articles tackling this topic and proposing different ideas of regulation. E.g., Brooklynn Armesto-Larson (2020) highlights that non-consensual pornography that includes deepfake pornography has an exterritorial character and local laws like The ENOUGH Act in the USA (the Ending Non-consensual Online User Graphic Harassment Act of 2017, that was not signed after all) are not sufficient.

Author hopes the cultural approach can help to understand deepfake pornography as not only a disgusting way of self-expression and the technology that penetrates personal borders and affects our trust in video information. (However, it does). But now we face the new toolkit's learning phase that shows potential to develop new possibilities for fans and for official media producers to create.